\documentclass{interspeech}

\usepackage{multirow}
\usepackage{xcolor}

\interspeechcameraready 

\title{Token-Level Logits Matter: A Closer Look at Speech Foundation Models for Ambiguous Emotion Recognition}

\author{Jule Valendo}{Halim}
\author{Siyi}{Wang}
\author{Hong}{Jia}
\author{Ting}{Dang}
\affiliation{}{University of Melbourne}{Australia}
\email{julevalendoh@student.unimelb.edu.au}

\usepackage{comment}

\begin{document}

\maketitle

\begin{abstract}
Emotional intelligence in conversational AI is crucial across domains like human-computer interaction. While numerous models have been developed, they often overlook the complexity and ambiguity inherent in human emotions. In the era of large speech foundation models (SFMs), understanding their capability in recognizing ambiguous emotions is essential for the development of next-generation emotion-aware models. This study examines the effectiveness of SFMs in ambiguous emotion recognition. We designed prompts for ambiguous emotion prediction and introduced two novel approaches to infer ambiguous emotion distributions: one analysing generated text responses and the other examining the internal processing of SFMs through token-level logits. Our findings suggest that while SFMs may not consistently generate accurate text responses for ambiguous emotions, they can interpret such emotions at the token level based on prior knowledge, demonstrating robustness across different prompts.

\end{abstract}

\section{Introduction}

Speech emotion recognition (SER) has experienced significant growth in areas such as mental health monitoring \cite{Elsayed2022} and human-computer interaction (HCI) \cite{Ramakrishnan2013}. Recent advancements in large language models (LLMs) have led to the development of advanced speech foundation models (SFMs), which integrate speech inputs with LLMs. These models incorporate both verbal content (e.g., text transcription) and vocal nuances (e.g., sighs, laughter, tones) in an end-to-end manner, offering a more comprehensive understanding of speech content and a sophisticated framework for SER~\cite{wu2024silentlettersamplifyingllms}.

Despite the promises of these SFMs for emotion recognition, they primarily focus on identifying single emotion classes, such as categorizing a sentence as either happy or sad~\cite{niu2024text, feng2024foundation}. However, human emotions are inherently multifaceted and ambiguous, influenced by cultural, contextual, and individual factors \cite{Mower2009}, which extend beyond simple classifications like happiness or sadness~\cite{wuemotion}. Yet, such emotion ambiguity is frequently neglected in automatic SER systems. This oversight mainly arises because the ground truth is set as a single label, based on majority voting from multiple annotations, which simplifies emotions into discrete categories~\cite{wu2024}. This simplification fails to capture the ambiguity of human emotions, limiting SER systems ability to understand human emotions and, consequently, hindering applications in HCI and mental health assessments, where discerning subtle emotional states is essential.

While a few studies have examined ambiguous emotion recognition, they predominantly utilize traditional modelling approaches such as Gaussian Mixture Regressions~\cite{dang2017investigation, dang2018dynamic}, probabilistic networks \cite{mohanty2020child}, multi-task neural networks \cite{zhang2018text}, Monte Carlo approach~\cite{wu2022novel} or neural ordinary differential equations~\cite{wu2024dual1}. Only one recent study explored LLMs for ambiguous emotion recognition~\cite{hong2024aerllmambiguityawareemotionrecognition}. However, this study converts speech inputs into textual descriptions of features and uses text-based LLMs, an approach that does not process speech directly and fails to capture nuanced emotional information in speech.

This study is the first to explore ambiguous emotion recognition using end-to-end SFMs built on top of LLMs. The aim is to deepen our understanding of how these models recognize ambiguous emotions. We propose two approaches: the first analyses text responses directly to evaluate how SFMs ``articulate'' ambiguous emotions. 
The second approach examines how these models interpret ambiguous emotions by analyzing the intermediate representations from SFMs, thereby reflecting their intrinsic ``conceptualization'' of emotions. We propose a token-level analysis in which the logits of emotion-related tokens within SFMs are extracted and processed to represent ambiguous emotions.

By comparing the ambiguous emotion extracted from the model’s text responses (``articulation'') with the token-level representation of its internal processing (``conceptualization''), our findings reveal that \emph{although SFMs may not consistently identify ambiguous emotions through articulation, they can more accurately process and recognize these ambiguous emotions at an token-level through posterior probability distributions}. Furthermore, the probability distributions are not sensitive to the prompts used, whereas the text responses are highly dependent on prompts. These insights highlight that pretrained SFMs possess the capabilities for recognizing ambiguous emotions, and our proposed token-level analysis provide an effective way to extract ambiguous emotion representation. This study opens up new avenues for enhancing chatbots or HCI applications through improved emotional intelligence in these systems.
\section{Related Work}

While the development of SER systems for single emotion classification has advanced for decades~\cite{LSTM1, YANG2024128177, fang2024multimodal, chandra2024}, the progress in recognizing ambiguous emotions is still lagging behind. Emotion complexity and ambiguity were first recognized in~\cite{mower2009interpreting}, which recommended classifying emotions based on soft-labels rather than relying solely on categorical hard labels. Subsequent studies have proposed using multiple classifiers to mimic individuals and simulate the ambiguity in emotions~\cite{zhou2022multi}, or treating ambiguous emotions as a completely different class, handling them as out-of-distribution data~\cite{wu2024handling}. A few other studies have suggested modelling emotions as distributions~\cite{Atcheson, dang2018dynamic, wu2022novel}. However, these approaches are generally based on traditional modelling methods. 

In the era of large-scale SFMs, employing a universal SFM trained on extensive speech data has opened up new possibilities for emotion understanding, moving away from reliance on task-specific small models. A recent study~\cite{huang2024emotionallynumbempatheticevaluating} explored the anthropomorphic capabilities of LLMs without fine-tuning, highlighting the capabilities of SFMs to capture emotional information from speech through expressive ``delivery skills'' inherently acquired during the training process. Another study~\cite{feng2024foundation} examined the capabilities of SFMs in recognizing single-label emotions with fine-tuning. However, these studies still overlook the ambiguity and complexity of human emotions. While~\cite{hong2024aerllmambiguityawareemotionrecognition} has shown that text-based LLMs can recognize ambiguous emotion to a certain extent, it does not explore the inherent capabilities of end-to-end SFMs processing speech directly.
\section{Methodology}
\subsection{Problem definition}
\begin{figure}[t!]
    \centering    \includegraphics[width=1\linewidth]{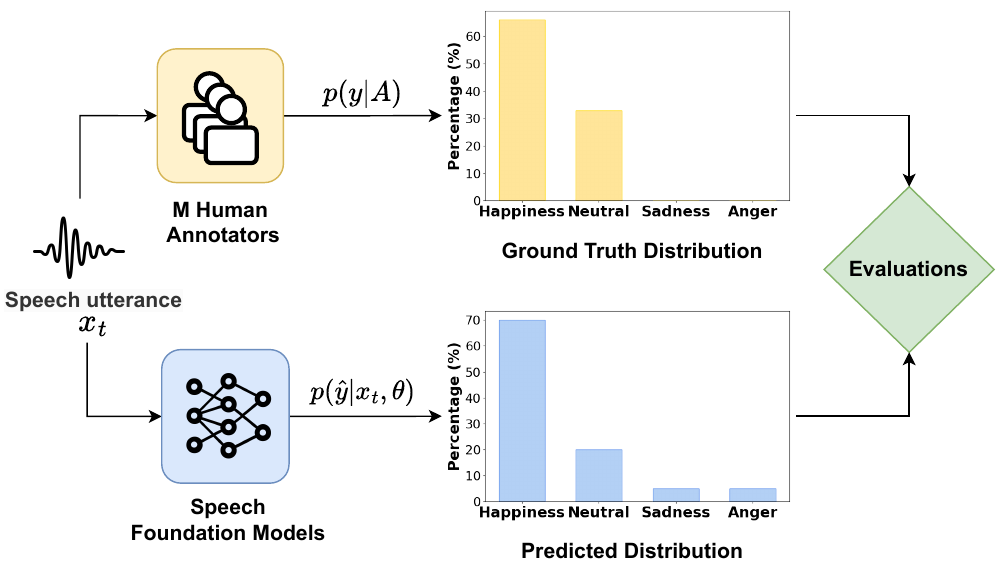}
    \vspace{-20pt}
    \caption{System overview. Speech utterances are processed by SFMs to generate emotion distributions, which are then compared with the ground truth inferred from $M$ human annotators. 
    }
    \vspace{-18pt}
    \label{fig:SystemOverview}
\end{figure}

As shown in Figure~\ref{fig:SystemOverview}, given a speech utterance $\bm x_t$, the objective is to infer the ambiguous emotion distribution $p(\hat{y}|\bm x_t, \bm \theta)$ using a SFM parameterized by $\bm \theta$. The predicted distribution encompasses $N$ emotion classes, with each probability indicating the likelihood of a specific emotion, while the overall distribution reflects the inherent ambiguity in emotional expression. This predicted distribution is directly compared with the ground truth distribution $p( y| A)$, derived from the annotations $A$ provided by $M$ human annotators.

\vspace{-5pt}
\subsection{System overview}\label{system overview}
As shown in Figure \ref{fig:LLMOutput}, we propose two approaches to predict the emotion distributions. 
The first approach focuses on text-level analysis that directly describe emotional ambiguity. The second approach conducts token-level analysis, extracting emotion distribution from the intermediate layers of SFMs. It examines how SFMs inherently process paralinguistic information in speech and the capabilities in emotion recognition they acquire during the pretraining paradigm. The prompt design will first be introduced, followed by two approaches for extracting emotion distributions.
\vspace{-5pt}
\subsection{Prompt design}\label{ambiguousPromptDesign}
To guide SFMs in understanding ambiguous emotions, the prompt should explicitly direct them to generate appropriate responses. Three key components have been considered:
\begin{itemize}
    \item \textit{Emotional distribution prediction}: The prompt requests the likelihood of each emotion being represented in the speech input, with probabilities expressed as percentages.
    
    \item \textit{Logical reasoning}: The model is directed to use logical reasoning to determine the output percentages.

\end{itemize}

\vspace{-10pt}
\begin{table}[h]
    \centering
    \begin{tabular}{p{0.45\textwidth}}
        \toprule
        \textit{Provide the likelihood (in percentages) that this audio represents each of the following emotions: anger, happiness, sadness, and neutral. Use logical reasoning to determine the percentages, but do not include this reasoning in your response.} \\
        \bottomrule
    \end{tabular}
\end{table}
\vspace{-10pt}

To ensure that the generated responses represent a valid distribution, where the probabilities in percentages sum to 100\%, additional constraints on the outputs will be applied.

\begin{itemize}
    \item \textit{Output constraints}:  If the generated responses for ambiguous emotion distributions fails to conform to a valid distribution, normalization is employed.
\end{itemize}

\vspace{-5pt}
\subsection{Extraction of emotion distribution}\label{MLLMResponseAmbiguous}

\begin{figure*}[ht]
    \centering  \includegraphics[width=1\linewidth]{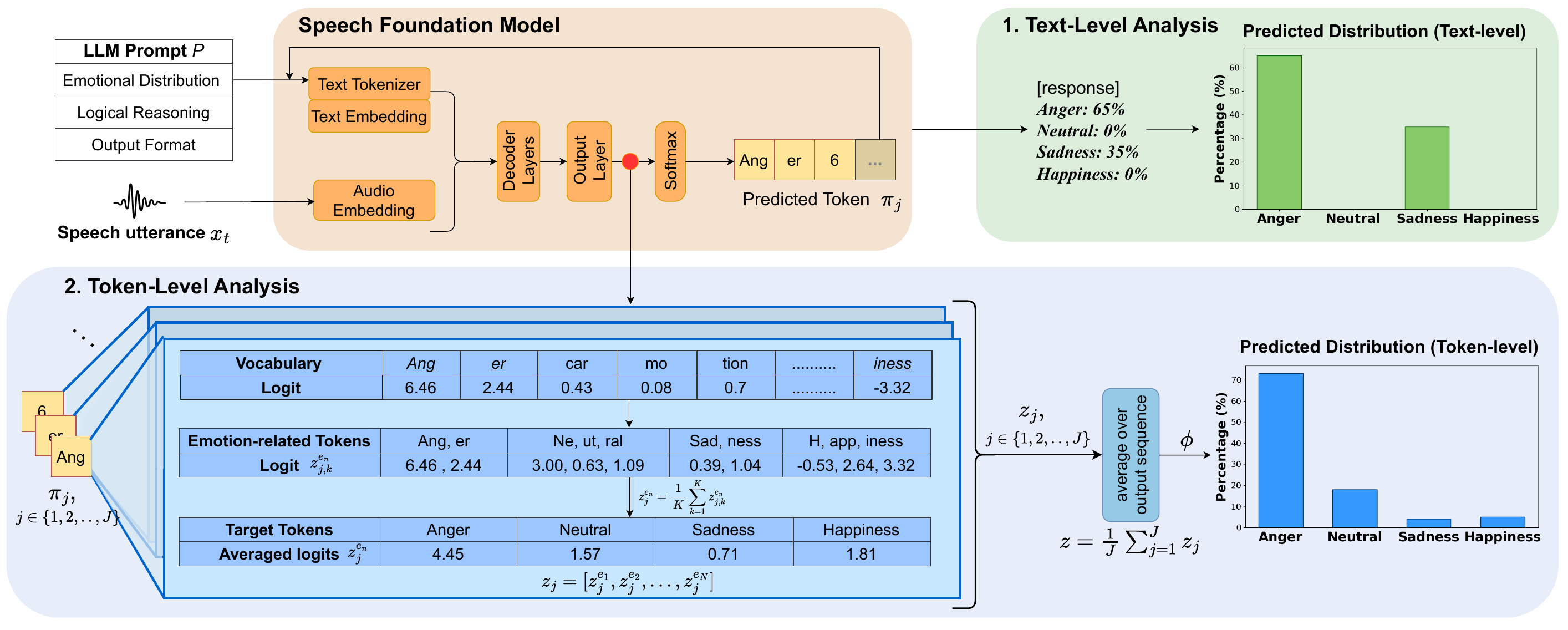}
  \vspace{-10pt}
    \caption{Framework for ambiguous emotion recognition using SFMs. By providing a prompt alongside speech, it enables the extraction of both the generated text and posterior probabilities at i) text-level and ii) token-level, respectively. }
    
    \vspace{-15pt}
    \label{fig:LLMOutput}
\end{figure*}

\subsubsection{Approach 1: Text-level analysis}
Given the speech utterances \(\bm{x}_t\) and the prompt \(P\), SFMs generate the text output describing the emotions (Figure~\ref{fig:LLMOutput}). Each of the emotions is associated with corresponding percentages.
A sample response given the prompt is as follows: Happiness: 0\%, Neutral: 0\%, Sadness: 35\%, Anger: 65\%, indicating that the SFM predominantly classifies the emotional content of the speech as anger, while also suggesting a probability of sadness. We convert the text responses into a discrete distribution %
and compare them with the ground truth distribution for evaluation.

\vspace{-5pt}
\subsubsection{Approach 2: Token-level analysis} \label{section3.4.2}

To further explore how SFMs inherently recognize ambiguous emotions, we focus on the token-level latent representations produced by the output layer of the SFMs, before they are converted into text output.  It should be noted that SFM generated the output tokens in an autoregressive manner. For example, given the speech utterances \(\bm{x}_t\) and the prompt \(P\), the SFM generates the first token ``ang'', followed by ``er'',``6'', then ``5'', and followed by ``\%'', until the complete text sequence was generated, represented by $[\pi_1, \pi_2, \ldots, \pi_J]$ where $\pi_j$ represents the $j^{th}$ output token. We extracted the logits $\bm z_j$ corresponding to each generated token $\pi_j$ sequentially, and aggregated them across the entire $J$ tokens to obtain the final logits representation to infer emotion distribution representations. 

We propose a two-step process: first, extracting the logits $\bm z_j$ from emotion-related tokens in the vocabulary, and then converting these logits into the final emotion distributions $\phi$.

\vspace{-10pt}
\paragraph*{Logit extraction from emotion-related tokens.}

As shown in Figure~\ref{fig:LLMOutput}, before generating each output token $\pi_j$ (e.g., ``Ang''), the inputs are first decoded to a latent vector that represents the vocabulary $\mathcal{V}$. This vector provides the probabilities for each token in the vocabulary, reflecting the likelihood of each being the next predicted token. We only extracted the logits corresponding to emotion-related tokens from the large vocabulary, e.g., [``ang'' ``er'' ``ne'' ``ut'' ``ral'', ``sad'' ``ness'', ``h'' ``app'' iness], 
represented as \(\bm z_j = [z_j^{ang}, z_j^{er}, z_j^{ne}, \ldots, z_j^{iness}]\).

As some emotion words are tokenized into multiple subword tokens, we average the logits across all subword tokens to obtain the logits for a single emotion word. For example, given ``Anger'' tokenized as ``Ang'' and ``er'', we average their corresponding logits \([z_j^{ang}, z_j^{er}]\). This process is applied similarly for other emotions and is represented as:
\vspace{-5pt}
\begin{equation}\label{eq:1}
    z_j^{e_n} = \frac{1}{K}\sum_{k=1}^K z_{j,k}^{e_n}
\end{equation}
where \(K\) is the number of tokens for each emotion word \(e_n\), and \(z_j^{e_n}\) is the averaged logits for the emotion \(e_n\). This leads to a vector \(\bm z_j = [z_j^{e_1}, z_j^{e_2}, \ldots, z_j^{e_N}]\) for each output token $\pi_j$.

Taking into account the entire output sequence, we average the logit representations \(\bm z_j\) across all output tokens \(\pi_j\). This captures the emotion distribution across the whole output sequence as \(\bm z = \frac{1}{J}\sum_{j=1}^J \bm z_j = [z^{e_1}, z^{e_2}, \ldots, z^{e_N}]\).

\vspace{-8pt}
\paragraph*{Conversion to Probabilities.}
To make the logits a valid emotion distribution, the second step converts the logits into probabilities by normalizing over $N$ emotions as follows:

\vspace{-5pt}
\begin{equation}\label{eq:3}
    \phi^{e_n} = \frac{z^{e_n}}{\sum_{n=1}^N z^{e_n}}
\end{equation}
resulting in the final emotion distribution.

These normalized probabilities offer insights into the model’s inherent perception of emotional content and its internal decision-making process.
\section{Experimental setup} \label{section:experimentalSetup}

\subsection{Dataset} 
The Interactive Emotional Dyadic Motion Capture (IEMOCAP) database \cite{busso2008iemocap} is used, consisting of approximately 12 hours of audiovisual recordings of dyadic conversations between actors. Each recording is annotated by three annotators each. Specifically, we select utterances that only include four emotion classes: Happiness, Anger, Sadness, and Neutral, resulting in 4,373 speech files. 

\vspace{-5pt}
\subsection{Implementation details}
\paragraph*{Models. }We adopt LTU-AS \cite{gong2023joint} as the SFM which leverages LLaMA architecture. It is open-sourced, allowing flexibility for extracting intermediate representations. LTU-AS demonstrated strong performance in understanding emotion, due to its capacity to encode non-verbal cues such as sighs and laughter which are critical for understanding emotional content. LTU-AS integrates a Whisper encoder-decoder architecture\cite{radford2022robustspeechrecognitionlargescale} with LLaMA. The Whisper encoder converts both spoken and non-spoken audio into embeddings. Non-spoken audio embeddings are projected into a format compatible with LLaMA, while spoken audio is transcribed by the Whisper decoder into text. This transcribed text is then converted into a distinct embedding, which is combined with the non-spoken audio embeddings and passed to LLaMA. Leveraging its pre-trained capabilities, LLaMA processes these embeddings, interprets emotional nuances, and generates a response that reflects emotional content from both verbal and non-verbal cues. Remote inference was performed through the HuggingFace Space platform, and local inference for posterior probabilities was conducted using two NVIDIA A100 80GB GPUs.

\vspace{-10pt}
\paragraph*{Evaluation. }
Two sets of evaluation metrics are used: ambiguity-based and accuracy-based. Ambiguity-based metrics aims to compare the predicted distributions and the ground truth distribution directly, including Bhattacharyya Distance (BD), Kullback-Leibler Divergence (KL), and the $R^2$ Score. 
Additionally, we evaluate the performance of single label prediction to the majority vote of the annotations, by selecting the most likely emotion from the predicted distribution. While the model is not optimized for single emotion recognition, this evaluation offers a basic understanding of how it perceives the dominant emotion. Accuracy and F1-score are used.
\section{Results}
\subsection{Performance on ambiguous SER}

Table \ref{tab:ambiguous_metrics} presents the performance of ambiguous SER using \emph{ambiguity-based metrics}, evaluating the entire distribution predictions. 2.1\% of the outputs produced invalid distributions and were omitted from the evaluation. Our approach at the token level achieves the best KL divergence and BD, outperforming the baseline that converts speech to text descriptions and uses text-based LLMs. 

More importantly, we found that the token-level analysis significantly outperforms the text-level analysis in recognizing ambiguous emotions, showing relative improvements of 51.71\%, 7.84\%, and 11.76\% in terms of KL divergence, BD, and $R^2$, respectively. This highlights that \emph{SFMs have the prior knowledge of ambiguity in emotional speech during the pretraining phase, but this ability does not fully translate into their text output.} The token-level analysis, therefore, provides an effective approach for inferring the nuances of ambiguous emotional expressions. 
While the baseline outperforms our approach in \(R^2\), it is possibly due to their use of the advanced Gemini-1.5 model, whereas we use LLaMA-based (7B) models. 

\begin{table}[t!]
  \caption{Performance for ambiguous SER. ($\uparrow$) means higher values are better, and ($\downarrow$) means lower values are better.}
  \vspace{-5pt}
  \label{tab:ambiguous_metrics}
  \centering
  \begin{tabular}{llccc}
    \toprule
    \textbf{Type} & \textbf{Method} & \textbf{KL} ($\downarrow$) & \textbf{BD} ($\downarrow$) & \textbf{$R^2$} ($\uparrow$) \\
    \midrule
    Baseline & Zero-Shot \cite{hong2024aerllmambiguityawareemotionrecognition} & - & 0.51 & \textbf{0.51} \\
    \midrule
    \multirow{2}{*}{Proposed} & Text & 2.05 & 0.51 & 0.34 \\
     & Token-level & \textbf{0.99} & \textbf{0.47} & 0.38 \\
    \bottomrule
  \end{tabular}
  \vspace{-5pt}
\end{table}

\begin{table}[t!]
  \caption{Impact of prompts for token-level analysis. }
  \vspace{-5pt}
  \label{tab:compare_single_ambiguous_metrics}
  \centering
  \small 
  \setlength{\tabcolsep}{6pt} 
  \begin{tabular}{llccc}
    \toprule
    \textbf{Type} & \textbf{Prompting} & \textbf{KL} ($\downarrow$) & \textbf{BD} ($\downarrow$) & \textbf{$R^2$} ($\uparrow$) \\
    \midrule
    \multirow{2}{*}{Token-level} & Ambiguous & 0.99 & 0.47 & 0.38 \\
                                 & Single    & \textbf{0.75} & \textbf{0.31} & \textbf{0.53} \\
    \bottomrule
  \end{tabular}
  \vspace{-10pt}
\end{table}

\begin{figure}[t]
    \centering
    \includegraphics[width=0.9\linewidth]{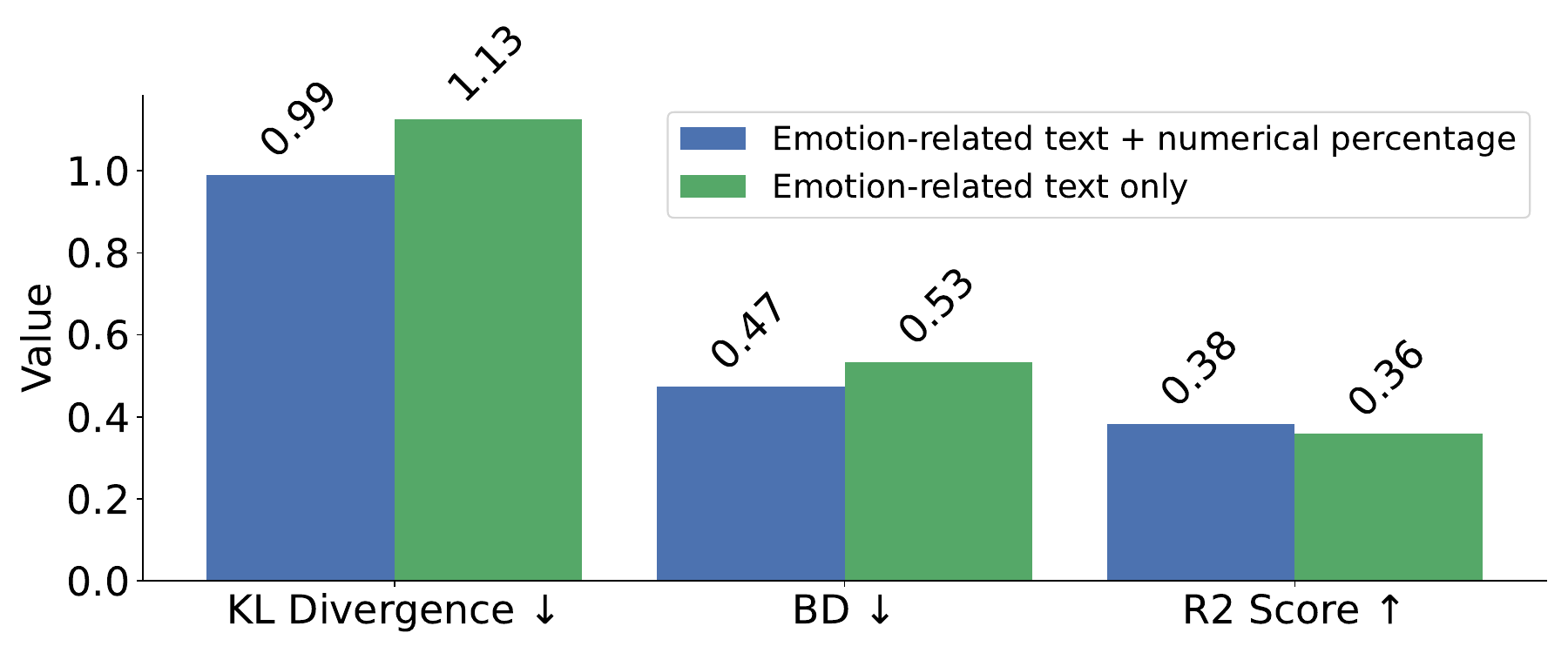}
    \vspace{-10pt}
    \caption{Performance comparison utilizing logits of i) both emotion-related text and numerical percentage and ii) only emotion-related text.}
    \label{fig:barplot}
\end{figure}
\vspace{-5pt}
\subsection{Impact of prompting at token level }\label{singleLabelPrompt}
To further investigate the robustness of token-level probabilities to variations in prompts for understanding emotional ambiguity, we employed prompts for single emotion recognition and compared them with prompts for ambiguous emotions. The single prompt we used is: ``\textit{You are an expert in identifying emotions from
speech. Predict the emotion of the audio from the choices [Happiness, Sadness,
Neutral, Angry]. Respond with only one of the emotion labels.''}. We also extracted the ambiguous emotion distributions at token-level for comparison. 

As shown in Table~\ref{tab:compare_single_ambiguous_metrics}, we found that: i) \emph{token-level analysis remains robust against different prompting strategies for understanding ambiguous emotions. Even with single prompting, the model can grasp the ambiguity.} ii) Single-emotion prompting demonstrates superior performance in understanding ambiguous emotions, outperforming ambiguous prompting across all metrics. This further confirms that SFMs have an inherent understanding of emotional ambiguity, remaining robust despite variations in the prompts. iii) The best performance achieved with single prompting surpasses the baseline in Table~\ref{tab:ambiguous_metrics}, highlighting the effectiveness of our token-level analysis approach for ambiguous emotion prediction.

\vspace{-5pt}
\subsection{Token-level distribution extraction comparison}\label{onlysinglelabel}
Instead of using the logits averaged across all output tokens \(\pi_j\) (section \ref{section3.4.2}), we also examined the logits averaged only across the emotion word tokens, i.e., a subset of \(\pi_j\) that ignores the tokens related to percentages. This aims to identify the optimal distribution extraction from logits, as shown in Figure \ref{fig:barplot}.

It is evident that incorporating logits from both the emotion-related text and the percentage outperforms using only the emotion-related text, suggesting that the logits for percentages also contain valuable emotional understanding. This is expected, as they reflect the model’s interpretation of the emotional categories. Moreover, SFMs operate in an autoregressive manner, where the percentage values capture meaningful information from the entire speech utterance as well as the previously generated text.

\subsection{Single emotion prediction}
To further understand how SFMs comprehend the dominant emotion, we infer the single emotion from our approaches and compared this with the ground truth majority vote emotion, as shown in Table~\ref{tab:acc_metrics}. We first evaluated our approach using a single prompt for single-emotion prediction and analyzed the text outputs for emotion recognition. This method achieved a superior accuracy of 58.06\%, outperforming other zero-shot learning approaches. These results highlight the advantages of leveraging speech input rather than relying solely on text input, as it enables more comprehensive processing of emotional traits from speech. It shows inferior performance compared to the LLM baselines with fine-tuning or few-shot learning, which is expected due to the additional training or the few-shot examples.

\begin{table}[t!]
\vspace{-8pt}
  \caption{Performance of single emotion recognition.}
  \vspace{-5pt}
  \label{tab:acc_metrics}
  \centering
  \small
  \setlength{\tabcolsep}{2pt} 
  \renewcommand{\arraystretch}{1.1} 
  \begin{tabular}{lccc}
    \toprule
    \textbf{Type} & \textbf{Method} & \textbf{Accuracy} & \textbf{F1-Score} \\
    \midrule
    \multirow{3}{*}{\textbf{Adaptation}} 
      & Fine-Tuned LLM \cite{gong2024listen} & \textbf{65.20} & -- \\
      & LLM Few-Shot \cite{hong2024aerllmambiguityawareemotionrecognition} & 58.75 & \textbf{0.59} \\
      \cmidrule{2-4}
      \multirow{2}{*}{\textbf{Zero-shot}} & LLM Zero-Shot \cite{hong2024aerllmambiguityawareemotionrecognition} & 48.12 & 0.49 \\
      &LLM Zero-Shot \cite{feng2023affect}&51.40&--\\
    \midrule
    \multirow{4}{*}{\textbf{Proposed}} 
      & Text (Single) & \underline{\textit{58.06}} & \underline{\textit{0.39}}  \\
      \cmidrule{2-4}
      & Text (Ambiguous) & 35.83 & 0.06 \\
      \cmidrule{2-4}
      & Token-Level (Ambiguous) & 52.64 & 0.31 \\
      & Token-Level (Single) & 55.90 & 0.30 \\
    \bottomrule
  \end{tabular}
  \vspace{-12pt}
\end{table}

Additionally, we inferred the most likely class from the predicted ambiguous distributions and compared it to the majority vote using both text-level and token-level analysis. It is important to note that the prompt for ambiguous emotions and the token-level analysis were not optimized for single emotion prediction. Nevertheless, the token-level analysis still achieved comparable or superior accuracy to zero-shot baselines. The lower F1-score is possible due to the inherent bias in SFMs towards the majority class. In summary, our approach not only captures the full emotion spectrum but also identifies the dominant emotion to a certain extent.
\vspace{-5pt}
\section{Conclusion}

Our study investigated the extent to which pretrained SFMs can interpret ambiguity in SER based on their prior knowledge and introduced two approaches to infer ambiguous emotion distributions at both the text and token levels. Our findings suggest that SFMs can recognize the nuanced ambiguity present in emotional speech due to its prior knowledge acquired during the pretraining phase, while this is not fully translated in their text-based outputs. The proposed token-level analysis offers an effective method for inferring both ambiguous emotion distributions and single emotions, outperforming state-of-the-art zero-shot baselines. These results highlight the potential of SFMs in applications where recognizing complex and ambiguous emotional states is crucial, such as mental health monitoring, and offer an innovative method to infer emotions from SFMs.

\bibliographystyle{IEEEtran}
\bibliography{mybib}

\end{document}